\documentclass[showpacs,showkeys,aps,pra]{revtex4}
\usepackage{amssymb}
\usepackage{amsfonts}
\usepackage{dcolumn}
\usepackage{amsmath}
\usepackage{graphicx}
\usepackage{psfrag,pstricks}
\usepackage{amsmath}
\usepackage{amssymb}

\def\eg{e.g.~}

\def\etc{ etc.~}

\begin{document}

\title{Chaos and Regularity in Semiconductor Microcavities}
\author{Hichem Eleuch$^{1,2}$}
\author{Awadhesh Prasad$^{1,3}$\footnote{Corresponding Author: Email- awadhesh@physics.du.ac.in,
Telephone +49 351 8712125, Fax +49 351 8711999.} }
\affiliation{$^1$Max Planck Institute for the Physics of Complex Systems,\\
N\"othnitzer Str. 38, 01187 Dresden, Germany \\
$^2$Institute for Quantum Science and Engineering, Texas A $\&$ M
University, College Station, Texas 77843, USA\\
$^3$Department of Physics and Astrophysics, University of Delhi, Delhi
110007, India}

\begin{abstract}
Our work presents a study on the nonlinear dynamical behavior for
a microcavity semiconductor containing a quantum well. Using an
external periodic perturbation in energy level we observe the
periodic-doubling, quasiperiodic, and direct route to chaos as
forcing strength is changed. For a particular case the riddled
basin for coexisting periodic and chaotic motions are observed.
These results suggest that the dynamics of exciton-photon is quite
complex in presence of external perturbation.
\end{abstract}


 \maketitle

\section{Introduction}

\bigskip The properties of the light emitted from a quantum well exciton
embedded in semi-conductor microcavities have drown attention of several
theoretical and experimental researchers in the last twenty years \cite%
{ex1,ex2,ex3,a4,a8,a11,aa1,aa2,a1,a2,a5}, given their potential application in
opto-electronic devices \cite{a6,a7}. Phenomena which are linked to quantum
electrodynamics, such as Rabi splitting and Autler-Townes doublets, have
been observed \cite{a8,a9,a10}.

\bigskip In the strong coupling regime when the coupling frequency between
the exciton and photon is larger than the relaxation frequencies
of the medium and the cavity, the vacuum Rabi splitting has been
observed. The degeneracy between the cavity resonance and the
medium is lifted and two lines appear in the intensity spectrum of
the exciton--cavity system. Semiconductor quantum wells exhibiting
narrow absorption lines corresponding to excitonic strengths of
these resonances make it possible to achieve the strong coupling
regime and to observe the vacuum Rabi splitting. In the strong
coupling regime also, modifications of quantum statistical
properties of the emitted light, were theoretical predicted and
experimentally proved
\cite{a11,a11b,a11c,a14,a15,a16,a17}. In 2004, the group
of Giacobino proved the existence of the optical bistability in
the semiconductor microcavities \cite{a18}. Bistability occurs
when there exist two possible stable states for the same set of
system parameters. Optical bistability can play a major role for
the conception of the optical memories paving the way to the
realization of the quantum computer.

Recently, multistability in semiconductor QED systems in the
strong coupling regime was observed in different experiments
\cite{a19,a20}. The multistability can arise due to different
effect such as polarization-dependent nonlinear interactions
\cite{a19} or thermal induced processes \cite%
{a21}.

Either in natural systems or in experimental realizations the
presence of an external forcing is unavoidable. Such external forcing
could be either noise (either from surroundings or inherent within
the experiment) or deterministic perturbation. The external
forcing in nonlinear dynamical systems has been extensively
studied. This forcing sometimes gives important and useful results
e.g.~ stochastic resonances \cite{reso,reso1}, chaos control
\cite{control,control1}, strange nonchaotic dynamics \cite{sna,sna1,sna2} etc.~ 
One aspect for exploring  the nonlinear behavior is to scan the parameters
space and observe how the dynamical complexity depends on the parameters
\cite{tabor,kaneko,ruelle,schuster}.
For a given set of parameters understanding the basins of coexisting attractors are also equally important,
particularly, when there are riddled basins \cite{riddle,riddle1,riddle2,riddle3}. 
The aim of this Letter is to analyze the dynamical behaviors 
(particularly routes to chaos \cite{tabor,kaneko,ruelle,schuster} and 
riddling \cite{riddle,riddle1,riddle2,riddle3}) of the
intra-cavity photonic intensity and the intensity of the
florescent light in the presence of periodic signals.

In this work we explore the dynamics of the field intensities for
high excitation regime inside a semiconductor microcavity
containing a quantum well. We observe that the periodic-doubling,
quasiperiodic and direct route chaos as forcing strength is being
changed. This shows the existence of various types of dynamics
over the forcing strengths. Furthermore coexisting periodic and
chaotic attractors are also observed with riddled basin.

The paper is organized as follows. In Sec. II, we review the
derivation of exciton-phonon interaction. This is followed in Sec.
III by results and discussion for three particular cases of
detuning where various complex dynamics are presented. Results are
concluded in Sec. IV.

\bigskip

\section{Model}

We consider a microcavity, containing a semiconductor quantum
well, with width of the order of the wavelength $\lambda$. The
quantum well is localized in a position, which corresponds to the
maximum of the electromagnetic field. This system is described in
detail in \cite{a22}.

Our discussions are limited to the semiconductors with two bands.
The electromagnetic field can excite an electron of the valence
band to the conduction band by creating a hole in the valence
band. The electron and the hole interact by giving excitonic
states, which are similar dependent states to the states of
hydrogen atoms. The state 1S which is the low bound state
(fundamental state) has the greatest oscillator strength. It is
for this reason that this processing of the interaction between
exciton - photon often takes into account only this state.
Neglecting the effects of the spins and the interaction with the
phonons we can write a Hamiltonian describing the system
\cite{a23,a24,a241}:

\begin{eqnarray}
H &=&\underset{k_{//}}{\sum }\hbar \omega
_{k,photon}a_{k_{//}}^{+}a_{k_{//}}+\underset{K_{//}}{\sum }\hbar \omega
_{K}b_{k_{//}}^{+}b_{k_{//}}+\underset{k_{//}}{\sum }\hbar g_{k_{//}}\left(
a_{k_{//}}^{+}b_{K_{//}}+b_{K_{//}}^{+}a_{k_{//}}\right) +  \notag \\
&&+\frac{1}{2}\underset{K_{//},K_{//}^{^{\prime }},q}{\sum }%
V_{q}^{eff}b_{K_{//}+q}^{+}b_{K_{//}^{^{\prime
}}-q}^{+}b_{k_{K//}}b_{K_{//}^{^{\prime }}}  \notag \\
&&-\underset{K_{//},K_{//}^{^{\prime }},q}{\sum \frac{\hbar g_{k_{//}}}{%
n_{sat}}}Aa_{k_{//}+q}^{+}b_{K_{//}^{^{\prime
}}-q}^{+}b_{k_{//}}b_{K_{//}^{^{\prime }}}+h.c.
\end{eqnarray}

The photonic and excitonic modes are quantified according to the
normal direction to the microcavity. The vectors of waves are thus
vectors parallel with the planes of the layers. In a perfect
planar cavity, the invariance by translation imposes that the
exciton with parallel wave vector can couple only with the same
photons with parallel wave vector $k_{//}=K_{//}.$

The first two terms of the Hamiltonian correspond to the proper energies of
photon and exciton, where $a_{k_{//}}$, $b_{K_{//}}$ are respectively the
annihilation operators of a photonic and excitonic mode verifying:

\begin{equation}
\left[ a_{k_{//}},a_{k_{//}^{^{\prime }}}^{+}\right] =\delta
_{k_{//}k_{//}^{^{\prime }}}
\end{equation}

\begin{equation}
\left[ b_{k_{//}},b_{k_{//}^{^{\prime }}}^{+}\right] =\delta
_{k_{//}k_{//}^{^{\prime }}}
\end{equation}

$\omega _{exc,K_{//}},$ $\omega _{ph,k_{//}}$ are the photonic
mode and excitonic mode frequencies of the cavity. The third term
corresponds to the exciton-photon coupling with a coupling
constant $g$. This constant is a function of the oscillator
strength by unit of area $\frac{f}{S}$, the effective length of
the cavity $L_{c}$, the dielectric permeability $\varepsilon $,
the mass $m_{0~}$and the free electron charge $e$ \cite{a27}:

\begin{equation}
g=\sqrt{\frac{~f}{2m_{0}~\varepsilon ~L_{c}~S}}
\end{equation}

and the fourth term represents the saturation of the interaction
photon-exciton due to the fermionic nature of the electrons and
the Holes, with \cite{a24,a241,a25,a26}:

\begin{equation}
n_{sat}=\frac{7}{16\pi \lambda _{x}^{2}}
\end{equation}

$\lambda _{x}~$being the radius of Bohr of the bidimensional exciton and it
is about $100\overset{o}{A\text{ }}$in the GaAs microcavity with a quantum
well.

The last term describes the interaction between exciton due to the
Coulomb interaction between components of an exciton and those of
a neighboring exciton. The effective potential is given by:

\begin{equation}
V_{q}^{eff}=V_{0}^{eff}\thickapprox \frac{6e^{2}}{\varepsilon A}\lambda _{x}
\end{equation}
where $A$ is the surface of quantification.We limit ourselves to the case of
a pumping having a normal incidence and which results in the excitation of
only one photonic mode of the cavity with wave vector $q=0$.

The momentum conservation imposes that this photonic mode is coupled only
with only one excitonic mode $K_{//}=0$. The Hamiltonian describing exciton
in the microcavity with a quantum well interacting with the photons is then:

\begin{eqnarray}
H &=&\hbar \omega _{ph}a^{+}a+\hbar \omega _{exc}b^{+}b+i \hbar g\left(
a^{+}b-b^{+}a\right) +\frac{1}{2}V_{0}b^{+}b^{+}bb  \notag \\
&&-\frac{\hbar g}{n_{sat}}\left( a^{+}b^{+}bb+b^{+}b^{+}ab\right).
\end{eqnarray}

If the excitation is very weak which correspond to very low
density, the distance which separates one exciton from its
neighbor is very large so that the interaction between excitons is
negligible as well as the effects of saturation and the exciton
can be considered as a pure boson and the exciton-photon
interaction can be modeled like an interaction between two
harmonic oscillators \cite{a28,a3b,a3b1}:

\begin{equation}
H_{lin}=\hbar \omega _{ph}a^{+}a+\hbar \omega _{exc}b^{+}b+\hbar g\left(
a^{+}b+b^{+}a\right).
\end{equation}

The third term of this Hamiltonian indicates the coupling between
exciton-photon and represent the exchange of excitations. The term $a^{+}b~$%
describes the creation of a photon and the annihilation of an exciton and
the term $ab^{+}~$describes the creation of an exciton and the annihilation
of a photon.The proper states of this system are called polariton states of
the cavity. The diagonalization of the Hamiltonian gives energies $E_{+}$and
$E_{\_}~$of polaritons in the cavity:

\begin{equation}
E_{\pm }=\frac{E_{cav}+E_{ex}}{2}\pm \frac{\hbar \Delta }{2}
\end{equation}

with

\begin{equation}
\Delta =\sqrt{\delta _{1}^{2}+4g^{2}}
\end{equation}

where the difference of frequency between frequencies of exciton and photon:

\begin{equation}
\delta _{1}=\omega _{cav}-\omega _{exc}.
\end{equation}

The corresponding polariton operators $P_{\pm }$ are linear combinations of
the excitonic and photonic operators:

\begin{equation}
P_{\_}=Xb-Ca
\end{equation}

\begin{equation}
P_{+}=Cb-Xa
\end{equation}

with X and C are the Hopfield coefficients:

\begin{eqnarray}
X &=&\frac{\delta _{1}+\Delta }{\sqrt{2\Delta \left( \delta _{1}+\Delta
\right) }}  \notag \\
C &=&\frac{2g}{\sqrt{2\Delta \left( \delta _{1}+\Delta \right) }}.
\end{eqnarray}

If we take into account the relaxation phenomena due to the finished
lifetime of excitons and photons that can be explained by the dissipations
of the cavity ($\gamma _{cav}$) and that of the exciton $\gamma _{exc}$.

We can modalize these effects by introducing the complex energies for the
exciton $E_{cav}-i\hbar \gamma _{cav}~$and for the photon of the cavity $%
E_{exc}-i\hbar \gamma _{exc}~\ $ \cite{a6}. Then we get complex
energies whose real part gives the energy of polariton and the
imaginary part represents the rates of polariton relaxation:

\begin{equation}
E_{\pm }=\frac{E_{cav}+E_{exc}}{2}-i\hbar \frac{\gamma _{cav}+\gamma _{exc}}{%
2}\pm \frac{\hbar \Delta }{2}
\end{equation}

\begin{equation}
\Delta ^{2}=\left( \delta _{1}-\hbar \left( \gamma _{cav}-\gamma
_{exc}\right) \right) ^{2}+4g^{2}.
\end{equation}

The polariton relaxation rates for the upper and lower branches are \cite%
{a28} :

\begin{eqnarray}
\gamma _{+} &=&\gamma _{exc}\cos ^{2}\left( \Theta \right) +\gamma
_{cav}\sin ^{2}\left( \Theta \right)  \notag \\
\gamma _{-} &=&\gamma _{exc}\sin ^{2}\left( \Theta \right) +\gamma
_{cav}\cos ^{2}\left( \Theta \right)
\end{eqnarray}

where $\Theta ~$is~the coupling angle defined by :

\begin{equation}
tg\left( 2\Theta \right) =\frac{-g}{2\left( \omega _{cav}-\omega
_{ex}\right) }=\frac{g}{2\delta _{1}}.
\end{equation}

In the case where the exciton and the cavity are in the resonance ($\omega
_{cav}=\omega _{ex})~$dissipations of the two branches are equal to the
average of excitonic and photonic dissipations:

\begin{equation}
\gamma _{+}=\gamma _{-}=\frac{\gamma _{exc}+\gamma _{cav}}{2}.
\end{equation}

If we take into account the effects of nonlinearity due to the excitonic
interactions and the effects of saturation we must reconsider the fourth and
fifth terms of the global Hamiltonian H.

Let us now derive the evolution of the excitonic and photonic fields in the
non linear regime where the density of exciton is high. The evolution of the
mean field operators in the interaction picture, by considering that the
fluctuations are weak compared to the average values, can be written as:

\begin{subequations}
\label{model}
\begin{align}
\frac{d\left\langle a\right\rangle }{d\tau } &=\varepsilon (t)+g\left\langle
b\right\rangle -\kappa \left\langle a\right\rangle -i\Delta _{a}\left\langle
a\right\rangle \\
\frac{d\left\langle b\right\rangle }{d\tau } &=-g\left\langle a\right\rangle
-\frac{\gamma }{2}\left\langle b\right\rangle -i\Delta _{b}\left\langle
b\right\rangle -2i\alpha \left\langle b^{+}\right\rangle \left\langle
b\right\rangle \left\langle b\right\rangle \\
\frac{d\left\langle a^{+}\right\rangle }{d\tau } &=\varepsilon
(t)+g\left\langle b^{+}\right\rangle -\kappa \left\langle a^{+}\right\rangle
+i\Delta _{a}\left\langle a^{+}\right\rangle \\
\frac{d\left\langle b^{+}\right\rangle }{d\tau } &=-g\left\langle
a^{+}\right\rangle -\frac{\gamma }{2}\left\langle b^{+}\right\rangle
+i\Delta _{b}\left\langle b^{+}\right\rangle +2i\alpha \left\langle
b^{+}\right\rangle \left\langle b^{+}\right\rangle \left\langle
b\right\rangle
\end{align}

where $\tau ~$is a dimensionless time normalized to the round trip time $%
\tau _{c}$ in the cavity$:$

\end{subequations}
\begin{equation*}
\tau =\frac{t}{\tau _{c}}
\end{equation*}

$\gamma ,~\kappa $ are the dimensionless decay rates of the exciton and
cavity photon:

\begin{equation*}
\gamma =\gamma _{ex~}\tau _{c};\kappa =\gamma _{ph~}\tau _{c}
\end{equation*}

the nonlinear coupling constant is normalized to $\frac{1}{\tau _{c}}:$

\begin{equation}
\alpha =\left( \frac{4g}{n_{sat}}X^{3}C+\frac{V_{0}}{\hbar }X^{4}\right)
\tau _{c}
\end{equation}

and $\Delta _{a},$ $\Delta _{b}$ are the dimensionless detuning

\begin{eqnarray}
\Delta _{a} &=&\left( \omega _{ph}-\omega _{L}\right) \tau _{c}  \notag \\
\Delta _{b} &=&\left( \omega _{ex}-\omega _{L}\right) \tau _{c}.
\end{eqnarray}

As this system is already 4-dimensional and hence the presence of
forcing makes the system more complex and difficult to understand
it analytically. Therefore the numerical simulation is used to
uncover the various dynamical states using RK4 integrator
\cite{nr}. We consider the step size $\Delta t=2\pi /5000$ for
integration. The dynamical studies are studied after removing
initial $10^{7}$ data points as transients. We explore here the
photon and exciton intensities $I_{a}=$ $\left\langle
a^{+}\right\rangle \left\langle a\right\rangle $ and
$I_{b}=\left\langle b^{+}\right\rangle \left\langle b\right\rangle
$ inside the cavity. As the fluorescent light is proportional to
the mean number of excitons ($I_{b}$), we are also
exploring the fluorescent light dynamics.

\section{Results and discussions}

There are various physical parameters in this system. We fix the
normalized parameters by $g=1.5$, $\kappa =0.12$, $\gamma =0.015$
. These parameters correspond to the experimental values
\cite{a18} in units of the inverse of the round-trip in the
microcavity. We concentrate our analysis to the strong pump field
where the non-linear phenomenon exhibits interesting
dynamics, in this case we choose a normalized amplitude of laser pump $%
\epsilon =200$ and the normalized excitonic interaction
coefficient $\alpha =0.00001$. At this set of parameters the
system has stable fixed point solution. However, particularly
$\epsilon $ is more prone to external influence. Therefore in this
study we consider this parameter only to observe the effect of
the external forcing. In this work we restrict our analysis to
the deterministic periodic forcing in $\epsilon $ i.e. $\epsilon
:\epsilon =1+f cos(\Omega t)$ where $\Omega$ is the perturbation frequency while $f$ is the strength of the 
forcing which is considered as bifurcation parameter \footnote{In this work we study only the case where the
perturbation frequency $\Omega=1$. 
However the transitions are similar for other values of $\Omega$ (\eg 1.5 and 2) as well as
for the case where the perturbation frequency is resonant to the Rabi frequency $\Omega=2g=3$.
 The other dynamics, such as strange nonchaotic attractor \cite{sna,sna1,sna2} due to irrational frequency, 
 will be reported latter.}. It is clear that for
$f=0$ the motion is fixed point solution. As we change the
magnitude of  $f$, various dynamical motions are
possible which we will discuss below for different values of the detuning $%
\Delta _{a}$ and $\Delta _{b}$. We limit our study to the case of the
resonance between the exciton and the cavity $\omega _{ex}=\omega _{ph}$. In
the strong regime where the values of the dissipation rates $\gamma $ and $%
\kappa $ are small compared to the coupling constant
exciton-photon $g,$ the systems exhibits two polaritonic branches
at $\omega _{ex}\pm g$ (see Eq. (15)). We explore the dynamics for
these two cases where the laser pump is resonant to one of the
polaritonic frequencies $\Delta _{a}=\Delta _{b}=\pm g.$ We also
analyze the case of the total resonance between the cavity, laser pump
and the exciton $\Delta _{a}=\Delta _{b}=0$.

\subsection{Case-I: .}

Let us begin with the case of total resonance $\Delta _{a}=\Delta _{b}=0$.
Shown in Fig. \ref{fig1}(a) is the spectrum of Lyapunov exponents (LE), $%
\lambda _{i}$ \cite{tabor} as function of forcing strength $f$.
The variation of LEs indicates the different dynamical states as
the forcing strength $f$ is changed. Here the zero Lyapunov
exponent is plotted as a dotted (blue) line while largest two
non-zero Lyapunov exponents are plotted as solid (black) and
dashed (red) lines. The figures in middle row represent the
intensity of the photons inside the cavity $I_{a}=$ $\left\langle
a^{+}\right\rangle \left\langle a\right\rangle $(black-solid line)
and the intracavity excitonic intensity $I_{b}=\left\langle
b^{+}\right\rangle \left\langle b\right\rangle $ (red-dashed line)
as function of time at different values of the forcing strengths (b)
$f=0.1$, (c) $f=0.804$ (d) $0.329$, and (e) $0.5$ which present the different
dynamical motions as period-one, period-six, quasiperiodic (QP) and chaotic
(C) respectively. These different dynamical motions can be further
confirmed by Poincar\'{e} section \cite{tabor} taken at $Re<a>=0$,
as shown in lowest row. The respective motions are the period-one
(single point), period-six (six points), quasiperiodic (closed curve) and chaotic
(scattered points). These results clearly demonstrate that there
are various possible dynamical states which depend on the forcing
strength.

As the forcing $f$ increased from zero, the motion goes from periodic (P)
to chaotic via QP. The left inset figure of Fig. \ref{fig1} shows
the expanded region in near to the transition marked as $F_{1}$.
It clearly indicates that the motion goes from periodic to chaotic
via quasiperiodic where two Lyapunov exponents are zero. This
transition can be termed as quasiperiodic route to chaos, a route
which has been found is generic in many other dissipative
dynamical systems \cite{kaneko,ruelle,schuster}.
Near to the transition $F_{2}$ there is an another route to
chaos where the periodic motion gets doubled and
leads to chaos as the forcing parameter is being increased. This is clearly shown in an expanded Fig. \ref{fig2} (a)
which demonstrates that whenever the largest non-zero Lyapunov exponent reaches to the zero value, the period gets doubled.
This period  doubling is further confirmed in Fig. \ref{fig2}  where $100$ consecutive maxima of $Re<a>$ 
(at a fixed parameter) is plotted as a function of  forcing parameter.
Here the periods get bifurcated  as periods $6\to 12 \to 24 ...$.  A representative period-six motion is
 shown in Figs. \ref{fig1}(c) and (g).
Apart from these transitions there are many other periodic windows, \eg $F_4$, $F_5$, $F_7$ 
\etc where one of the above routes occurs.

However as the forcing strength $f$ \ is decreased near the marked
transition $F_{3},$ the motion becomes chaotic from periodic where there is
sudden jump in Lyapunov exponents. This suggests that there is direct
transition from periodic to chaotic as shown in inset figure near to the
transition $F_{3}$. This suggest that this is an another type of route to
chaos. Here due to high dimensionality of the system it is difficult to find
the normal form to understand the exact type of bifurcation. Apart from the
above routes/dynamics there are various periodic windows of different
periods also e.g.~ near to $F_{4}$. These results indicate that in the
presence of the external forcing this system exhibits various type of dynamics.

\subsection{Case-II: $\Delta_a=\Delta_b=g$}

This situation presents the case where the laser pump is resonant to the
upper polaritonic branch. Similar to the case-I, we plot in Fig. \ref{fig3}
(a) the largest two Lyapunov exponents, $\lambda _{i}$ as a function of the
forcing strength $f$. The figures in middle row show the intensities $I_{a}$
(black-solid line) and $I_{b}$ (red-dashed line) as a function of time at
different forcing strengths (b) $f=0.06,(c)0.18$ and (d) $0.8$ for periodic,
quasiperiodic, and chaotic attractors respectively. These different
dynamical motions can be further confirmed by Poincar\'{e} section taken at $%
Re<a>=0$, as shown in lowest row, for respective motions which are the
period-one (single point), quasiperiodic (closed curve) and chaotic
(scattered points) motions. These clearly demonstrate that these various
dynamical states also possible in this case as well.

Here as the forcing strength is increased from zero, the period motion becomes
quasiperiodic (which is in wider region than that of the Case-I)
and then is chaotic near to the transition $F_{1}$. Here also the
transition $F_{1}$ where motion goes from periodic to chaotic is
direct in the sense that the periodic motion suddenly becomes
chaotic. This transition, similar to $F_{3}$ transition in Case-I,
is still not exactly clear due to inability to get its normal
form.

\subsection{Case-III: $\Delta_a=\Delta_b=-g$}

This situation presents the case where we pump resonantly to the
lower polaritonic branch. In Fig. \ref{fig4} (a) are plotted the
spectrum of Lyapunov exponents, $\lambda _{i}$ as a function of
the forcing strength $f$. Similar to Figs. \ref{fig1} and \ref{fig3}
shown in middle row are the intensities $I_{a}$ (black-solid line)
and $I_{b}$ (red-dashed line) as function of time at different
forcing strengths (b) $f=0.5$ and (d) $1.7728$
for periodic and chaotic attractors respectively. The corresponding Poincar%
\'{e} sections, taken at $Re<a>=0$, are shown in lowest row, which confirm
the period-one (single point), and chaotic (scattered points) motions
respectively. These clearly demonstrate that, in this case as well, various
dynamical states are also possible.

However the transition from periodic to chaotic and then to
periodic is quite different than from the previous Cases-I and II.
Here we don't find neither period-doubling nor quasiperiodic nor
direct route to chaos as observed in previous cases. A remarkable
feature of this case is that there appears to be a parameter
interval, near the transition $F_{1}$ in Fig. \ref{fig4}(a),
where the system exhibits wild fluctuations as a function of the
forcing strength $f$. These fluctuations persist on small scales,
as shown in a blowup of marked box in inset figure. Such
fluctuations typically indicate co-existing attractors with
complicated basins \cite{riddle,riddle1,riddle2,riddle3} because of a small change in the
parameter $f$, it can lead to a completely different attractor.
Here we do observed coexisting periodic (Fig. \ref{fig5}(a)) and
chaotic (Fig. \ref{fig5}(b)) motions. The basin of these
multistable attractors are shown Fig. \ref{fig5}(c) with
different initial conditions corresponding to periodic (open
circles) and chaotic (star) motions. In the diagonal region,
interwoven initial conditions show the presence of complicated
basins where both of the attractors coexist. In order to confirm
the complicated basins, it is plotted in Fig. \ref{fig5}(c) the
basin of expanded region of marked box in (d). This demonstrate
that basin is riddled \cite{riddle,riddle1,riddle2,riddle3} i.e.~ if an initial condition
goes to periodic attractor then for any slight change in the $f$
parameter, say within the change $\delta $, it may go to chaotic
attractor, and vice-versa. However, this phenomenon of riddling
can be verified via the computation of an uncertainty exponent.
Fixing a perturbation, $\delta $, and randomly choosing a pair of
systems within a region, the parameters are termed uncertain if
they yield different asymptotic states. The fraction of uncertain
parameter pairs, $q(\delta )$, typically decreases with $\delta $
as a power

\begin{equation*}
q(\delta) \propto \delta^\beta
\end{equation*}

which defines the uncertainty exponent $\beta $ \cite{riddle,riddle1,riddle2,riddle3}. Results are plotted in Fig. %
\ref{fig5}(e) where the exponent is approximately zero (the best
fit to the data yields $\beta =0.0008\pm 0.0004$) which is typical
of riddle or riddle-like basins \cite{riddle,riddle1,riddle2,riddle3}. The practical
implication is that in this region, the attractor cannot be
predicted no matter how small the uncertainty in the specification
of parameters. This suggests that this system exhibits complicated
dynamics which not only depends on initial conditions but also on
the perturbation parameter.

\section{Conclusion}

In summary, we have studied the dynamical behavior of the photonic and
excitonic intensities inside a semiconductor microcavity with a quantum
well. As the fluorescent light is proportional to the intracavity excitonic
intensity, we have explored also the fluorescent light dynamics.

We have observed first that by varying the external perturbation we can
change the intracavity field intensities. By increasing the amplitude of the
external perturbation $f$ and independently of the resonance configurations
mainly the photon intensity inside the cavity is increased.

By pumping the upper polaritonic branch (caseII) there are more
excitons than photons inside the cavity and by pumping the lower
polaritonic branch (case III) we have the inverse situation (see
Figs. 2 and 3). This can be interpreted by the fact that excitonic
nonlinearity induces more excitons in the upper branch than in the
lower one. The enhancement of the photon intensity is more
pronounced by pumping resonantly to the upper polaritonic branch.

A modification of the trajectory regularity is observed by
increasing\ the external forcing strength  $f$. As shown in Fig
1,2 and 3, by increasing the perturbation amplitude $f$ the
dynamics of the system changes from stable point (Fig. 1a, Fig. 2a,
Fig. 3a) to chaotic motion (Fig. 1c, Fig. 2c and Fig. 3c). The
quasiperiodic motion is totally absent in  the case III.
Moreover the basin of attractor in III are quite complicated.

The Case-III shows different behavior in comparison to I and II.
The chaos is observed in wide range of parameters for I and II.
From this we can conclude that the regularity of the dynamics for
the polaritonic lower branch is less sensitive to the external
perturbation. Furthermore we have observed that Hyper chaotic
behavior (more that one Lyapunov exponents are rather that zero)
in I but not in II and III. This means that by pumping in the
middle between the two polaritonic branches the system presents
very complex dynamics. In addition, we can deduce that the
perturbation affects less the polaritonic lower branch than the
upper branch( see Figs. 2 and 3). By pumping with a laser frequency
detuned from the polaritonic frequencies the system exhibits high
sensitivity to the perturbations. This effect can originate from
the nonlinear effects induced by the exciton-exciton interaction.
The nonlinearity affects more the frequency ranges detuned from
the polaritonic frequencies and induces high dependence of the
system to the external perturbations.

\section*{Acknowledgements}

The authors thank E. Siminos for valuable comments and acknowledge
the financial support and the hospitality of MPIPKS.

\newpage

{\bf{Figure captions:}}

\vskip1cm

Fig.1 {(Color online) Case-I: $\Delta_a=\Delta_b=0$. (a) The two largest
Lyapunov Exponents with function of forcing strength $f$. Middle row:
intensities $I_a $ (black-solid line) and $I_b$ (red-dashed line) as a
function of time at different values of the forcing strengths (b) $f=0.1$ (period-one
motion) (c) $f=0.804$ (period-six motion) (d) $f=0.329$ (quasiperiodic motion) 
and (e) $f=0.5$ (chaotic motion). The scale of y-axises in b), c), d), and e) are the same.
 The (f), (g), (h) and (i) are the
 Poincar\'e section at $Re<a>=0$ in the intensities space $I_a-I_b$ for b), c) d) and e)
 respectively. The dotted line
in a) represents the zero Lyapunov Exponent. The inset figures are the
expanded regions near to different transitions as indicated by arrows.}

\begin{figure}[t]
\includegraphics[width=5in]{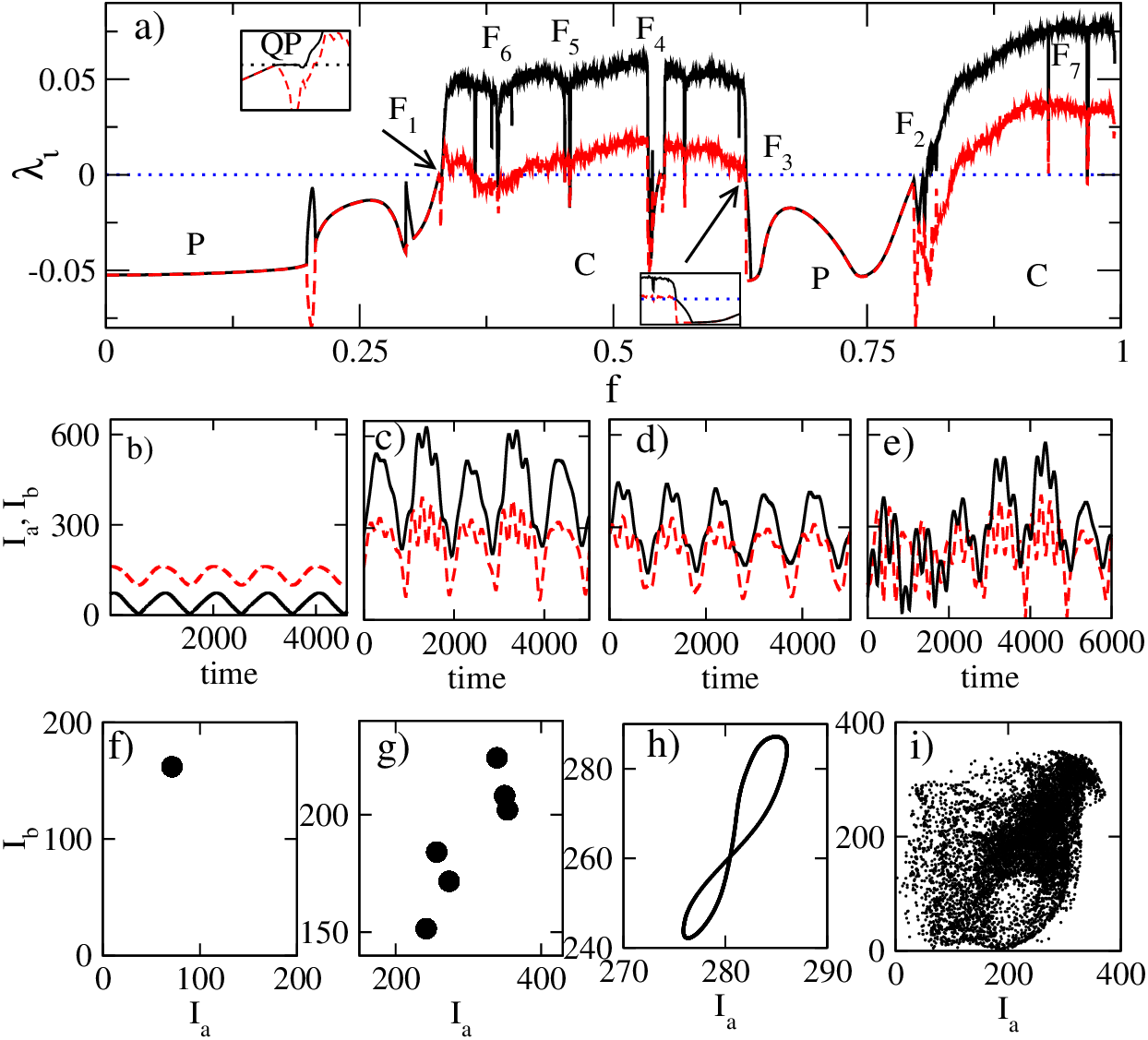}
\caption{}
\label{fig1}
\end{figure}

\begin{figure}[t]
\includegraphics[width=5in]{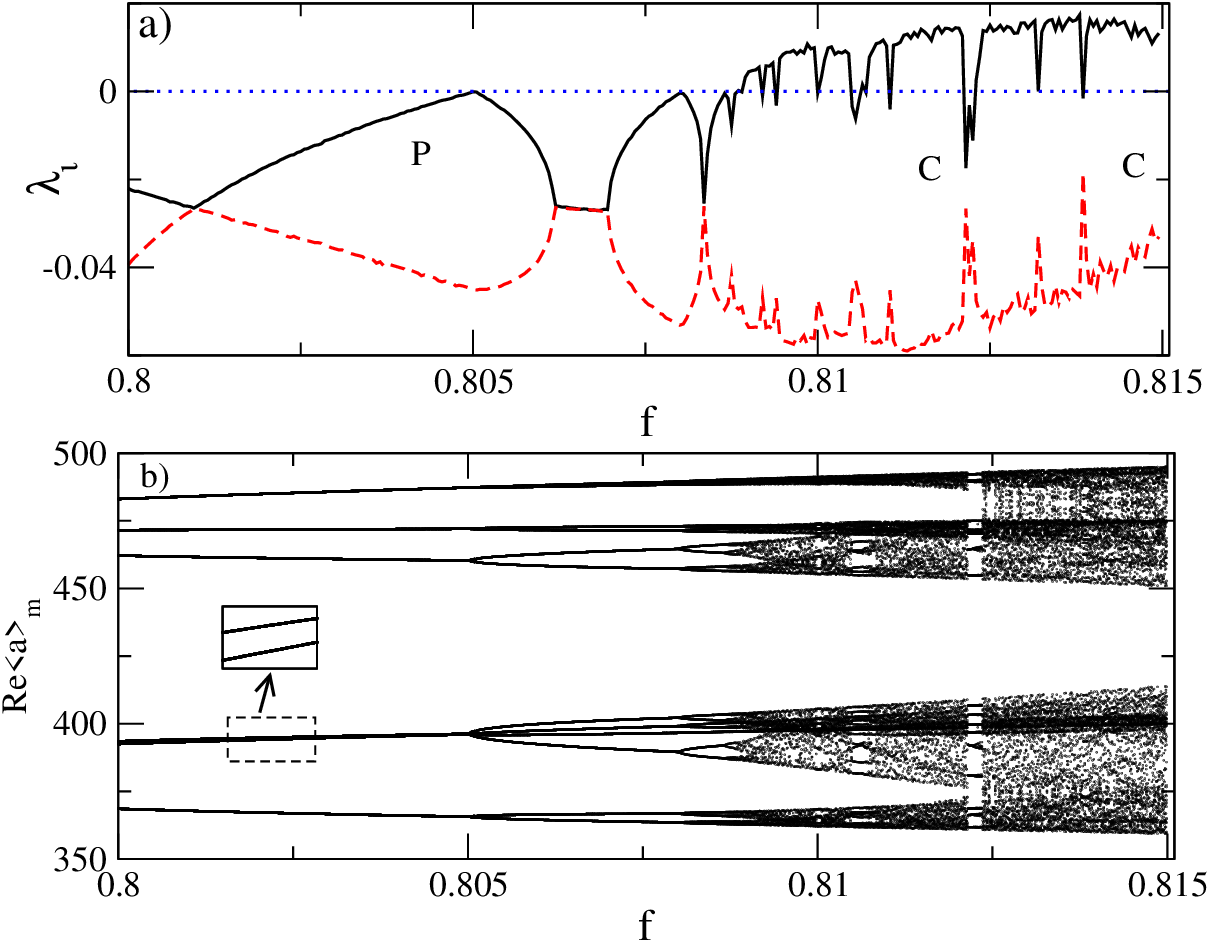}
\caption{}
\label{fig2}
\end{figure}

\begin{figure}[t]
\includegraphics[width=5in]{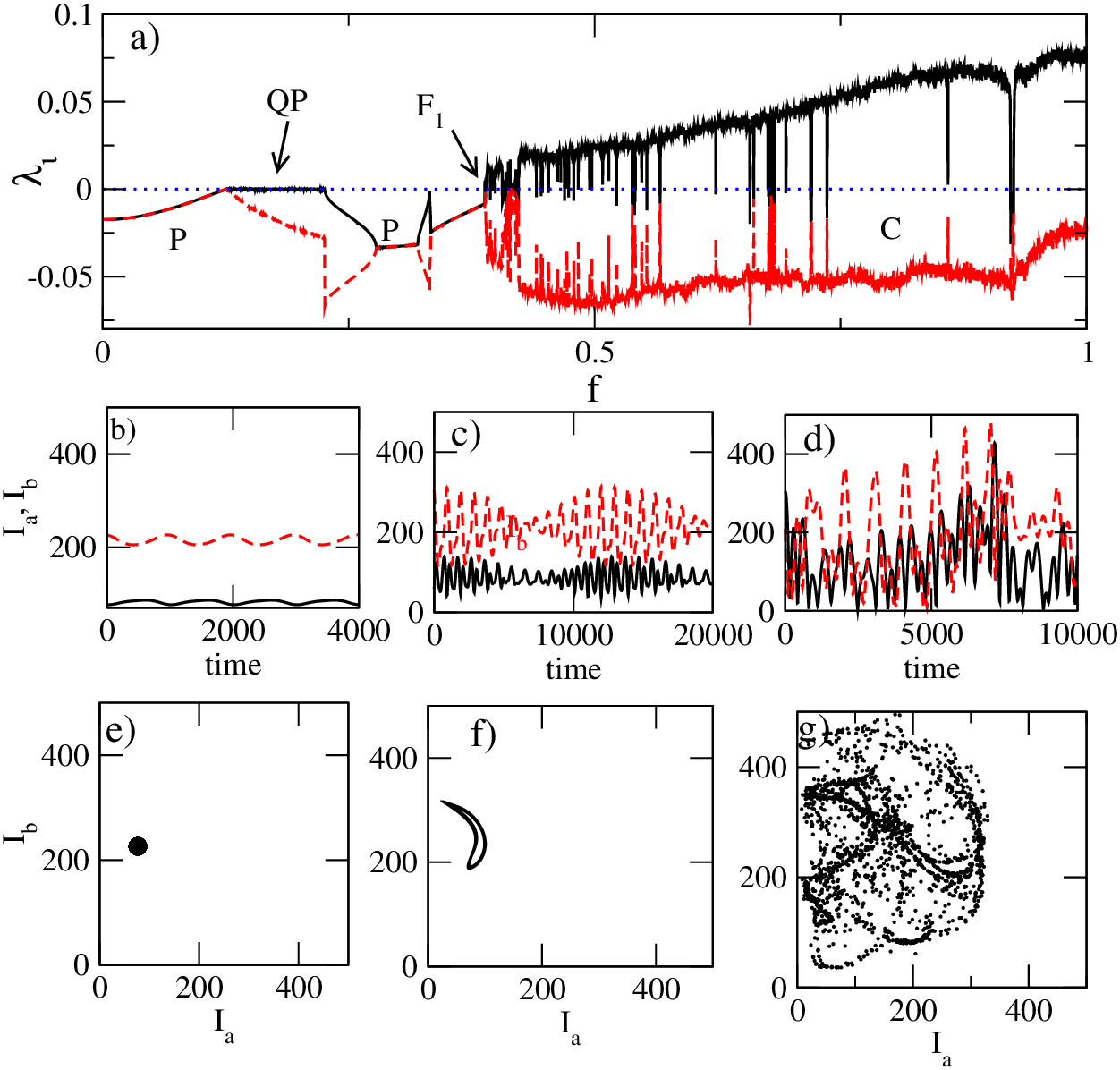}
\caption{}
\label{fig3}
\end{figure}

\begin{figure}[t]
\includegraphics[width=5in]{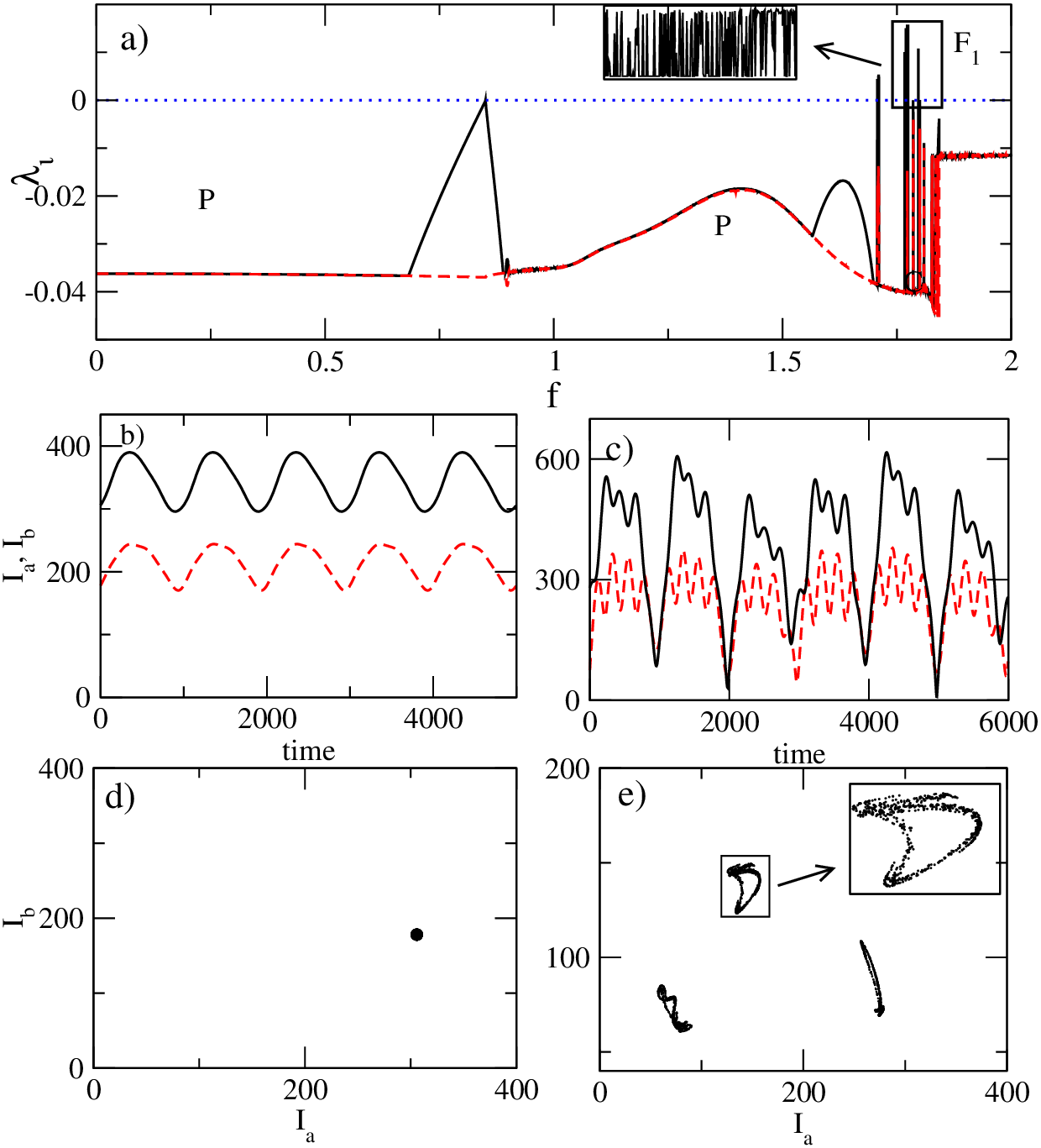}
\caption{}
\label{fig4}
\end{figure}

\begin{figure}[t]
\includegraphics[width=5in]{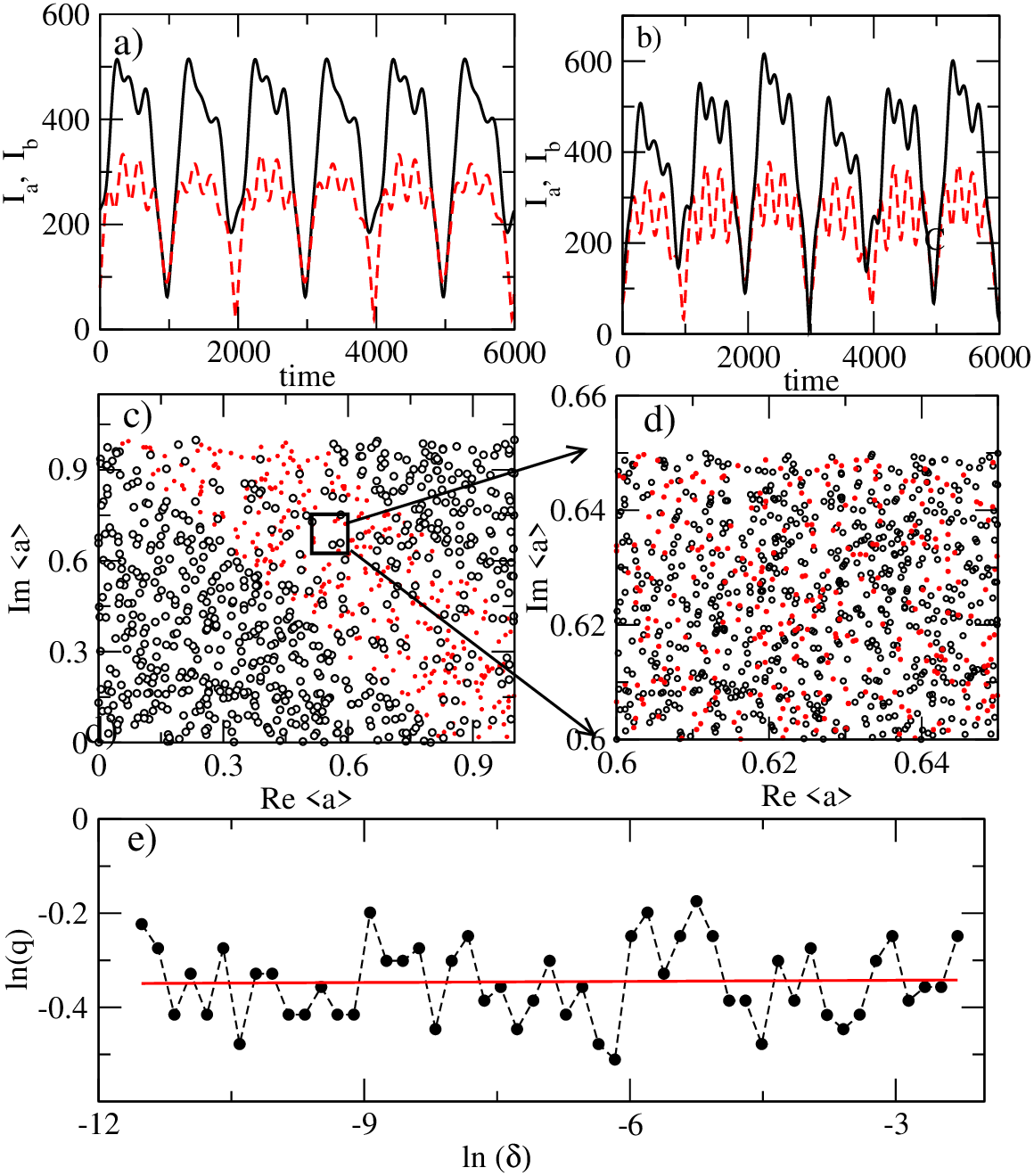}
\caption{}
\label{fig5}
\end{figure}

\vskip1cm
Fig.2 { (Color online) (a) The two largest
Lyapunov Exponents  and (b) the bifurcation diagram, $Re<a>_m$ (maxima of $Re<a>$) as a function of the forcing
 strength $f$ near to the marked transition $F_2$ of Fig. \ref{fig1}(a).
The inset figure in b) is expanded region of dotted box.}

\vskip1cm
Fig.3 {(Color online) Case-II: $\Delta_a=\Delta_b=g.$ The two largest
Lyapunov Exponents as a  function of the forcing strength $f$. Middle row:
intensities $I_a$ (black-solid line) and $I_b$ (red-dashed line) as a
function of time at different forcing strengths (b) $f=0.0.06$ (periodic
motion) (c) $f=0.18$ (quasiperiodic motion) and (d) $f=0.8$ (chaotic
motion). The (e), (f) and (g) are the Poincar\'e section at $Re<a>=0$ in
intensities space $I_a-I_b$ for b), c) and d) respectively. The dotted line
in a) represents the zero Lyapunov Exponent.}

\vskip1cm
Fig.4 {(Color online) Case-III: $\Delta_a=\Delta_b=-g$. The largest two
Lyapunov Exponents as a function of the forcing strength $f$. Middle row: $I_a$
(black-solid line) and $I_b$ (red-dashed line) as a function of time at
forcing strengths (b) $f=0.5$ (periodic motion) and (c) $f=1.7728$ (chaotic
motion). The (d) and (e) are the Poincar\'e section at $Re<a>=0$ in
intensity space $I_a-I_b$ for b) and c) respectively. The dashed line in a)
corresponds to the zero Lyapunov Exponents. The inset figures in a) and e)
are the expanded regions of marked boxes in the respective figures with more
data points.}

\vskip1cm
Fig 5 {(Color online) Case-III: $\Delta_a=\Delta_b=-g$. The presence of
coexisting attractors (a) period and (b) chaotic as function of time at
a forcing strength $f=1.77273$. (c) Then basin of these coexisting attractors in $Re <a>
$ vs $Im <a>$ with fixed $Re<b>=Im<b>=0$. (d) is an expended basins of
marked box in c). (e) shows the fraction of uncertain parameter pairs (out
of a sample of 300) as a function of the parameter perturbation $\protect\delta
$ (see text for detail).}

\newpage

\newpage

\newpage

\newpage

\newpage

\end{document}